\documentclass[conference, 10pt]{IEEEtran}
\IEEEoverridecommandlockouts
\usepackage{cite}
\usepackage{algorithmic}
\usepackage{graphicx}
\usepackage{textcomp}
\usepackage{xcolor}
\usepackage{color, colortbl}
\usepackage{epsfig}
\usepackage{epstopdf}
\usepackage{mciteplus}
\usepackage{stfloats}
\usepackage{amsmath}
\usepackage{amsfonts}
\usepackage{amssymb}
\usepackage{url}
\usepackage{array}
\usepackage{setspace}
\usepackage{paralist}
\usepackage{suffix}
\usepackage{mathtools}
\usepackage{psfrag}
\usepackage{multirow}
\usepackage{tabularx}
\usepackage[T1]{fontenc}
\usepackage{balance}
\definecolor{Gray}{gray}{0.9}
\def\BibTeX{{\rm B\kern-.05em{\sc i\kern-.025em b}\kern-.08em
    T\kern-.1667em\lower.7ex\hbox{E}\kern-.125emX}}
\begin{document}

\title{Energy Data Visualizations on Smartphones for Triggering Behavioral Change: Novel Vs. Conventional}

\makeatletter
\newcommand{\linebreakand}{%
  \end{@IEEEauthorhalign}
  \hfill\mbox{}\par
  \mbox{}\hfill\begin{@IEEEauthorhalign}
}
\makeatother

\author{
\IEEEauthorblockN{Ayman Al-Kababji, Abdullah Alsalemi, Yassine Himeur, Faycal Bensaali}
\IEEEauthorblockA{\textit{Dept. of Electrical Engineering} \\
\textit{Qatar University}\\
Doha, Qatar \\
\{aa1405810, a.alsalemi, yassine.himeur, f.bensaali\}@qu.edu.qa}
\and
\IEEEauthorblockN{Abbes Amira}
\IEEEauthorblockA{\textit{Institute of Artificial Intelligence} \\
\textit{De Montfort University}\\
Leicester, UK \\
abbes.amira@dmu.ac.uk}
\linebreakand
\IEEEauthorblockN{Rachael Fernandez, Noora Fetais}
\IEEEauthorblockA{\textit{Dept. Of Computer Science \& Engineering} \\
\textit{Qatar University}\\
Doha, Qatar \\
\{rf1405233, n.almarri\}@qu.edu.qa}
}
\maketitle
\begin{abstract}
This paper conveys the importance of using suitable data visualizations for electrical energy consumption and the effect it carries on reducing said consumption. Data visualization tools construct an important pillar in energy micro-moments, i.e., the concept of providing the right information at the right time in the right way for a specific power consumer. Such behavioral change can be triggered with the help of good recommendations and suitable visualizations to convey the right message. A questionnaire is built as a mobile application to evaluate different groups of conventional and novel visualizations. Conventional charts are restricted to bar, line and stacked area charts, while novel visualizations contain heatmap, spiral and appliance-level stacked bar charts. Significant findings gathered from participants' responses indicate that they are slightly inclined towards conventional charts. However, their understanding of the novel charts is better by 8\% when the analysis questions are investigated. Finally, a question is answered on whether a group of visualizations should be discarded completely, or some modifications can be applied.
\end{abstract}


\section{Introduction}\label{sec1}
It is needless to say how dependent people's lives have become on the existence of electric energy. Whether it is in houses, offices, gardens or even the streets, electric energy can be found everywhere. It exists in industrial, residential and commercial fields where it is used for similar and dissimilar kinds of loads. However, people's exploitation of such energy has become excessive, and the world is witnessing a surge with the spread of the coronoavirus (COVID-19) due to people's need to work from home for social distancing \cite{BBC2020}. Not only that, but the environmental effect of producing electric energy is also greatly neglected by the public.

According to the International Energy Agency, in 1973, a worldwide final energy consumption was estimated to be 4,659 Mtoe/year, where 439 Mtoe (9.41\%) of this energy was consumed as electricity \cite{Agency2019}. While in 2017, final energy consumption doubled to become 9,717 Mtoe/year, and with it doubled the electric energy consumption percentage to become 18.9\% (1,838 Mtoe) \cite{Agency2019}. This represents a world-wide  rise in “electrical” energy consumption estimated around 319\% in the last four and a half decades only. The number is expected to increase more in the future due to the increase in earth population and the direct link between energy consumption and high-quality of life. This growth is significant and is accompanied by excessive CO\textsubscript{2} emissions due to the dependence on non-renewable energy sources being the main source of electric energy.

Raising awareness among the public regarding energy consumption is crucial, and many researchers and governments are working towards it. The aim is to educate people and provide them with enough information to understand their energy consumption levels. Consequently, fostering sustainable behavioral change in which they decide to reduce their consumption. In this context, an important term arises, which is the ``micro-moment'' concept. It is defined by small time-based events in which the end-user conducts an energy action (e.g. switch on/off an appliance, enter/leave a room, adjust the settings of a device) \cite{alsalemi2019ieeesystems}. These moments can help in raising awareness about energy efficiency and can stimulate behavioral change. 
Behavioral change moments can be triggered through providing consumers with relevant significant information regarding their consumption; thus, comes the importance of using data visualization tools. Graphs and charts are the essential tools to convey such information in a concise and informative way to draw subjective conclusions. These graphs, if built properly, can influence consumers to be more energy efficient and more environmentally friendly. This can work regardless of the incentive behind their behavioral change, be it environmental, economical, societal, etc. 

According to the authors in \cite{Herrmann2018}, the type of employed data visualization affects power consumers' understanding of residential energy consumption. In addition, they conclude that time-series data visualizations are not preferred and opt the use of summary overview visualizations. Lastly, they recommend the visualization of dis-aggregated energy at the appliance level, but they have not studied this visualization's contribution in lowering energy consumption. In \cite{spangher2019engineering}, the direction is towards using ambient-type visualizations instead of engineering-type visualizations burdened with numbers the public are not familiar with. The study focuses on how to relate the reduction in consumption to the end-result, such as environmental benefit, to trigger behavioral change. For \cite{Watanabe6598515}, they opt to use personalized kind of charts for each consumer based on what he/she favors to see or visualize. Personalized visualizations are based on bar, line and pie charts. On the other hand, authors in \cite{Costanza2012} went with allowing consumers to make their own visualization using a tool named FigureEnergy instead of generating it for them as in \cite{Watanabe6598515}. Lastly, the authors from the EnergyAware framework create a mobile application that shows historical and real-time energy consumption with an alerting system for overloading scenarios \cite{buono2019energyaware}.
To the authors' best knowledge, conducting a questionnaire using a mobile application has never been done before. Moreover, the investigated data visualizations with the highest ratings in this study will be used later in the (EM)\textsuperscript{3} framework deployment.

This study is a part of a greater framework, namely the (EM)\textsuperscript{3} framework \cite{alsalemi2019ieeesystems, alsalemi_access_2020} depicted in Fig. \ref{fig:em3}. It is a research initiative targeting to induce domestic energy-saving behavior with the use of artificial intelligence and data visualization. One of its significant pillars is the data visualization tools deployed, which signifies the chances that the recommendations suggested by the recommender system are accepted.

\begin{figure}[!ht]
\centering
\includegraphics[trim={0.5in 0.5in 0.5in 0.6in},clip, width=0.95\linewidth]{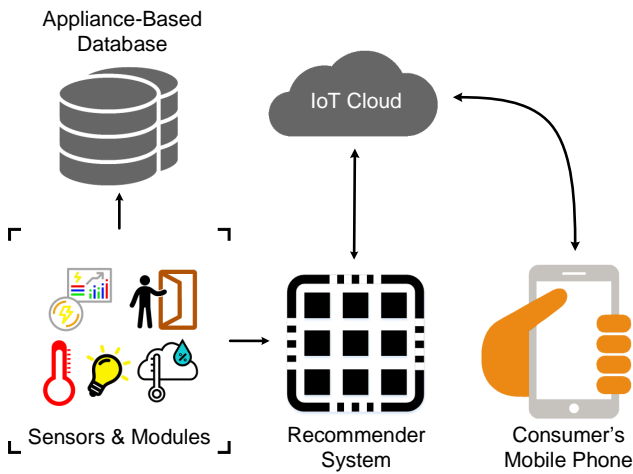} 
\caption{Overview of the (EM)\textsuperscript{3} framework}
\vspace{-0.5em}
\label{fig:em3}
\end{figure}

The objective of this paper is to find informative, easy-to-understand and aesthetically appealing charts to be used as data visualization tools. Thus, a questionnaire, embedded in a mobile application, is built to choose which charts are favored among the public. It is done by gathering participants' responses to conventional and novel charts, and based on the findings, a decision is made on which is more suitable for the public. 

To that end, the remainder of this paper is organized as follows. Section \ref{sec2} highlights the adopted study scheme and the tested visualization plots. Section \ref{sec3} delves into the implementation of the visualizations and the questionnaire on smartphones. Section \ref{sec4} presents responses summary and focuses on some of the findings. Finally, the paper is concluded in Section \ref{sec5} where the future work is also highlighted.

\section{Data Visualization Study Implementation}\label{sec2}

The study is structured to comprise two groups of visualizations, namely conventional and novel. Each group includes three charts for each participant to contemplate. Conventional charts are the type of charts used heavily in many fields, including the energy field (e.g. line, bar, scatter). Novel charts, on the other hand, are the new charts created/utilized in this study to be used specifically in the field of energy. Both chart groups are shown in Fig. \ref{fig:plots}.

The process that participants go through from downloading the application until answering all relevant questions is highlighted in Fig. \ref{fig:participation-process}.

\begin{figure}[!ht]
\centering
\includegraphics[width=1\linewidth, trim={0.6in 0.9in 0.95in 0.6in}, clip] {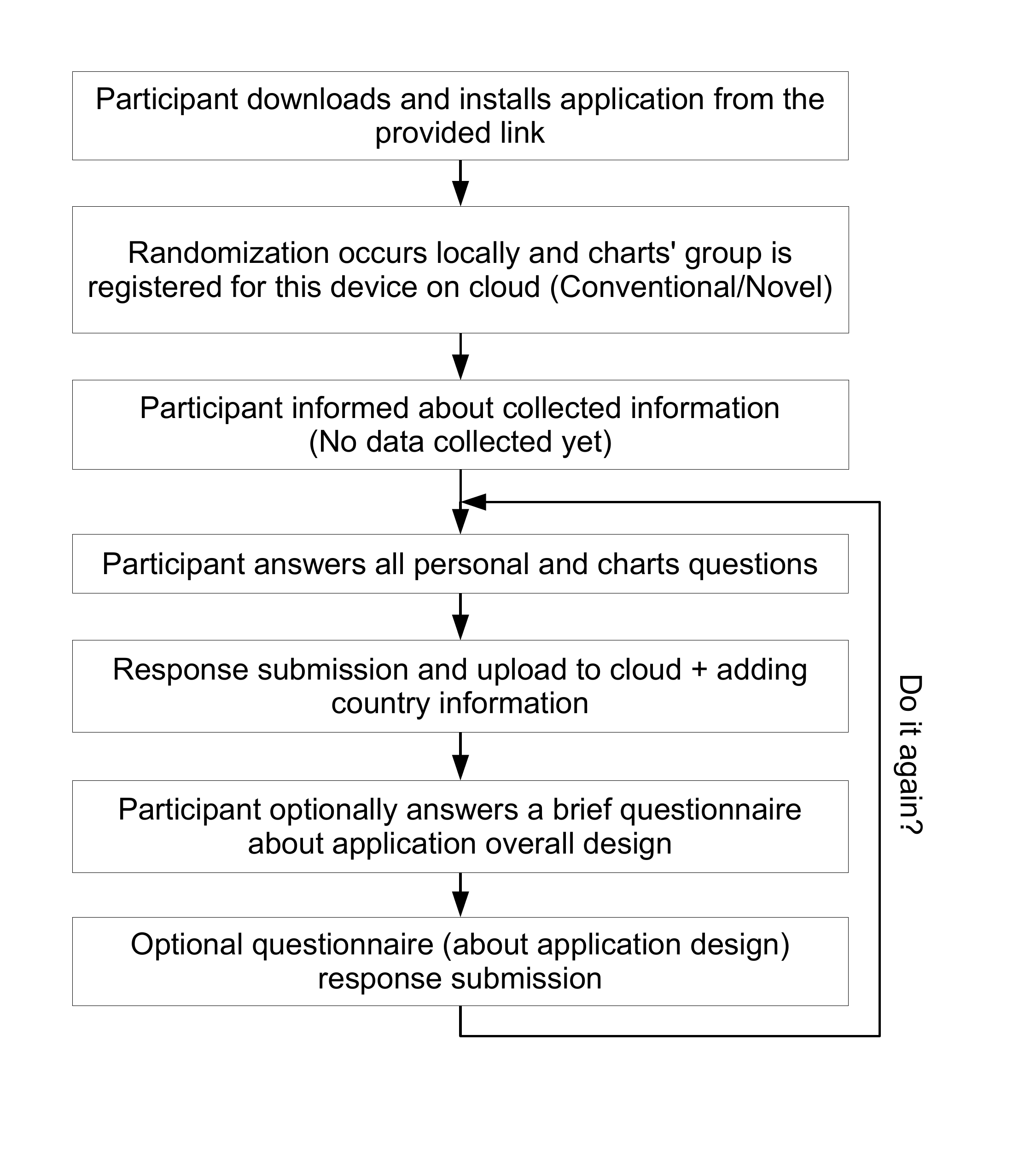}
\caption{Participation process}
\vspace{-0.5em}
\label{fig:participation-process}
\end{figure}

\begin{figure*}[!ht]
\centering
\includegraphics[trim={0.1in 0.2in 0.1in 0.2in},clip,width=1\linewidth]{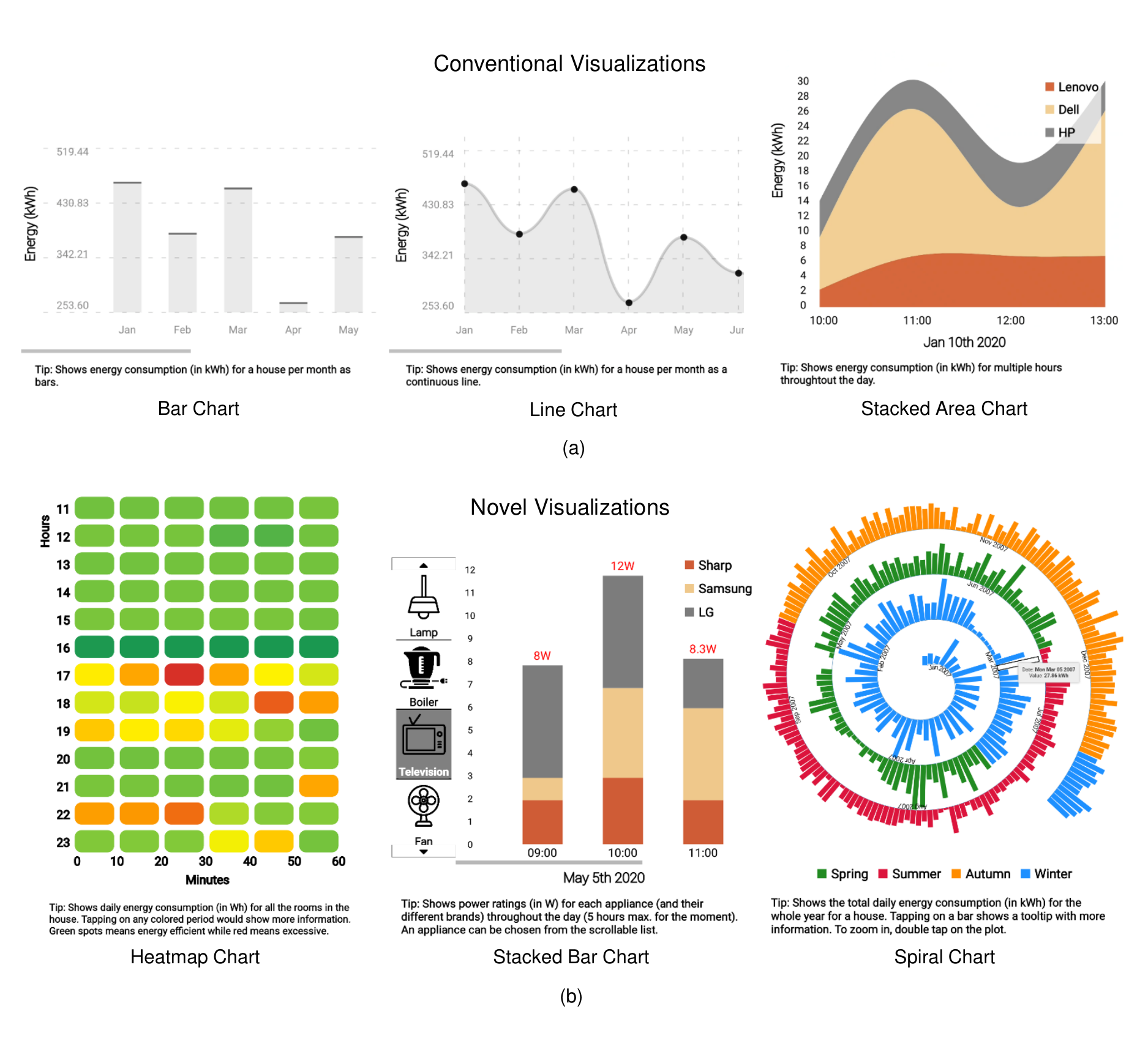} 
\caption{(a) Conventional and (b) novel visualizations created for the study}
\vspace{-1.5em}
\label{fig:plots} 
\end{figure*}

To be part of the study, participants download and install an application through a distributed link (sent through emails, social media channels and (EM)\textsuperscript{3} project website\footnote{http://em3.qu.edu.qa/index.php/data-visualization-app}). Once the participant initiates the application, the type of charts (i.e., conventional or novel) that he/she sees is locked in. In other words, the participant does not know about the existence of the other group and is questioned on one type of these charts (either conventional or novel).

At the beginning, the participant inserts their personal information. It is worth mentioning that participants are well-informed about the collected data, including the country they currently reside in using their IP addresses. The personal information includes gender, age, participant's profession and if he/she had any experience related to the field of data visualizations. This information can be utilized to group participants based on their age group, their background and/or their expertise in the area of data visualizations. Once inserted, the participant will answer questions related to each chart. Each chart has seven questions, divided into four qualitative and three quantitative (analysis) questions. Qualitative questions are the same for all charts, and they are mentioned here:

\begin{enumerate}
\item How effective is the provided visualization in portraying power consumption information? \textit{(1 not effective at all – 5 very effective)}
\item How easy to understand is the provided visualization in portraying power consumption information? \textit{(1 not easy at all – 5 very easy)}
\item How visually pleasing is the provided visualization? (\textit{1 not visually pleasing at all – 5 very visually pleasing)}
\item In terms of quantity, how do you describe the amount of data presented? \textit{(options: a. sparse (not enough), b. adequate (enough), c. excessive (very complex))}\footnote{Remaining quantitative questions per chart can be found in this link: http://em3.qu.edu.qa/wp-content/uploads/2020/05/em3-data-vis-appendix-A.pdf}
\end{enumerate}

After completing the questionnaire, the participant is commended for his/her contribution and faces an optional questionnaire to answer regarding the application design. The participant has the chance to even restart the questionnaire and let someone else do it on their smartphone.

The deployed charts used in this study are shown in Fig. \ref{fig:plots}. Some of the charts such as the line and bar charts show the aggregated energy consumption data for a whole month. Spiral and heatmap show the data aggregated on a daily basis and on 10 minutes period respectively. Even though the stacked area and stacked bar charts energy consumption is aggregated, the actual energy consumption per appliance can still be observed. The stacked bar chart has an extra advantage, which is choosing the desired appliance from a list.

Real energy data are portrayed to add authenticity and value to the charts where participants will be observing actual recorded data from existing datasets. The ``Individual household electric power consumption Data Set'' (abbreviated as IHEPCDS) dataset contains daily energy consumption in Watt-hour (Wh) where a single sample is collected per minute. This dataset is created from metering a household in France between December 2006 and November 2010. For visualization purposes, the values sub\_metering\_1, sub\_metering\_2 and sub\_metering\_3 are assumed to be rooms within that household. They are aggregated in different manners to be used in the bar, line, spiral and heatmap charts. Both bar and line charts show the energy consumption of year 2007 from IHEPCDS dataset, aggregated in monthly periods \cite{french_dataset_2012}. For the spiral, the energy consumption data of year 2007 are aggregated as days to show the consumption throughout the year. The heatmap aggregates daily energy consumption into periods of 10 minutes. The date (Sunday, April 1\textsuperscript{st}, 2007) is purposefully chosen for the heatmap as it is the day with the highest variations within that year; hence, richer data can be visualized. For the stacked area and stacked bar chart, the created data are inspired from the ACS-F2 dataset \cite{acs-f2-dataset} but the values from this dataset are not used as is.

\section{Mobile Platform Design}\label{sec3}
In this section, the design behind the application that contains the questionnaire is discussed. The scheme is based on having 3 screens that each has its own purpose. For illustration purposes, the application usage cycle is summarized in Fig. \ref{fig:applicationUsageCycle}.

\begin{figure}[!ht]
\centering
\begin{center}
\includegraphics[width=0.75\linewidth, trim={2.5in 1in 2.5in 1in}]{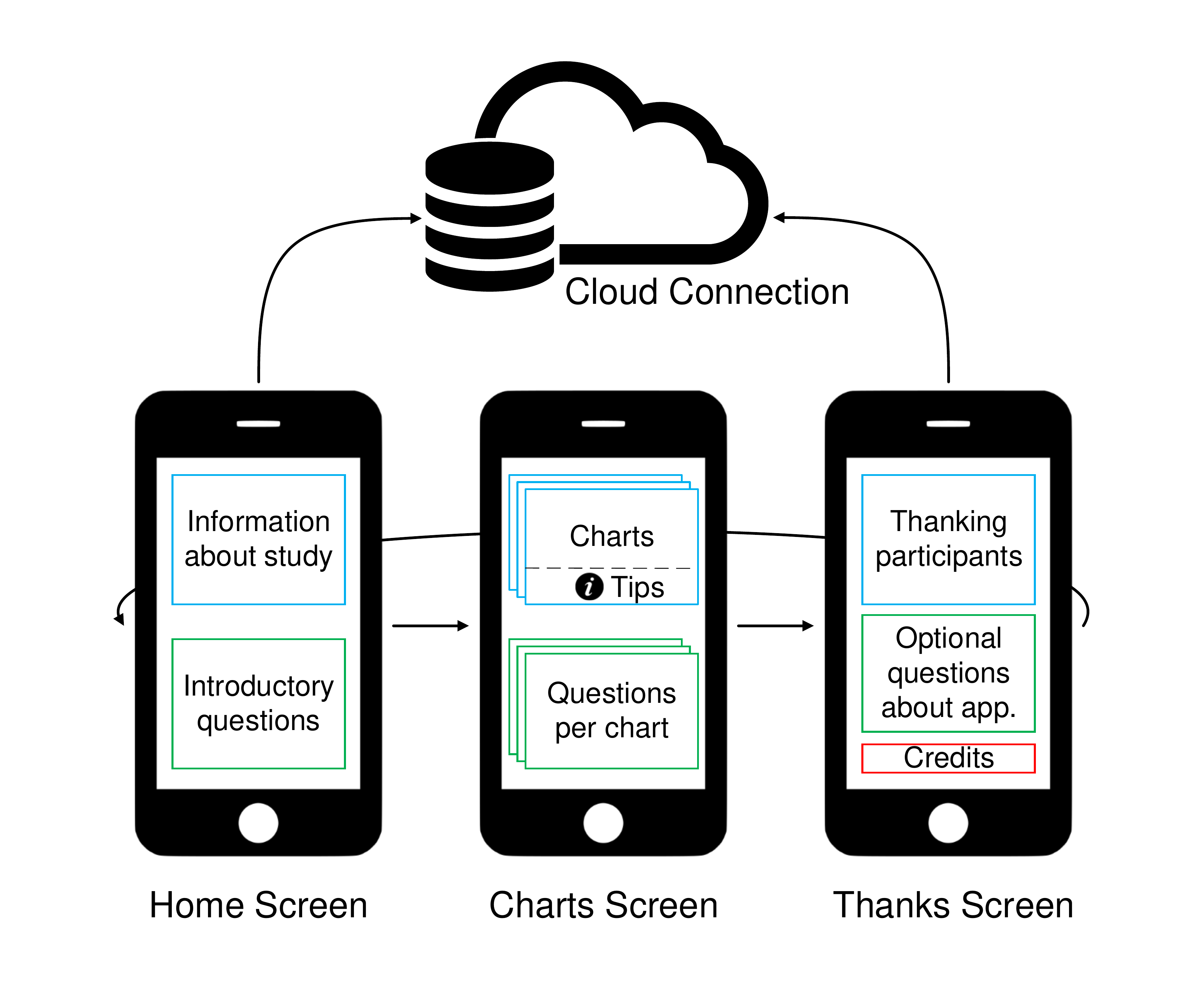}
\caption{Application usage cycle}
\vspace{-2em}
\label{fig:applicationUsageCycle}
\end{center}
\end{figure}

The Home Screen serves as an introductory screen to inform the participants briefly about the conducted research, the purpose of the study, what kind of information is collected, and to what aim they are used. If the participant proceeds with the application, a consent to collect the data is implicitly granted. Moreover, few questions are also shown in the first screen to determine the demography that the participant belongs to and if he/she has any experience in data visualization. At this stage, the type of charts that the participant will see is determined for the rest of the study by random selection. Next, Charts Screen shows the three different charts from the same charts group (conventional/novel), and collects the participants' answers to each one of them. Finally, Thanks Screen concludes the charts' related questionnaire and thanks the participants for their time. On a side note, an optional questionnaire is introduced in this screen to quantify participants' feedback regarding the flow and design of the application itself, and the duration spent to solve the questionnaire. It includes participants' observation section where he/she can comment on anything regarding the charts or the application. In the end, a credits section to acknowledge copyright-material is also presented. Since uploading the participants' answers requires the existence of internet connection, a listener function for internet connection status is activated within the application, where it prevents participants from completing the questionnaire in the absence of internet connection.

The application is developed using React Native, a cross-platform framework for building native applications for iOS, Android and even web applications. The significance of this framework is that the developer needs to build a single, universal script, which significantly reduces development time. JavaScript (JS) and its syntax extension JSX are the main tools to fully utilize React Native, which enables developers to use HTML-like tags within JS files. One of the interesting features is its capability to be used in an existing native application (iOS or Android) or in creating a whole application from scratch \cite{react-native-website}. The latter scenario is considered for this study to accelerate development time.

Since the focus of this study is not data storage and management, a ready-to-use cloud database (DB) is incorporated to store participants' responses in a single destination and to facilitate access for analysis later on. Firestore, which is a DB service provided by Firebase, is a cloud-hosted, No-Structured Query Language (NoSQL) DB that can be accessed directly on almost any platform through a secure application programming interface (API). Its data model is utilized by inserting responses as documents containing fields that map to participants' answers. These documents are stored in a collection that behave as a container for the created documents \cite{firebase-firestore-website}. This model is suitable since the desired behavior is storing responses incoming from the same smartphone to a designated node on the cloud. Another important feature Firebase provides for any of its DBs is the ability to establish ``Rules'' that enable certain users to read/write to the database. In a sense, it provides security against eavesdroppers from reading, writing or deleting any of the collected information.

\section{Results and Discussion}\label{sec4}
In this section, a discussion of the acquired questionnaire results is conducted. Table \ref{table:study-distribution} shows a brief summary of the participants' distribution, noting that these experts are self-proclaimed.

\vspace{-1em}
\begin{table}[htbp]
\caption{Study Distribution}
\begin{center}
\begin{tabular}{|>{\bfseries}c|c|c|c|}
\hline
\textbf{Type}&\textbf{Conventional}&\textbf{Novel}&\textbf{Total} \\ \hline
Non-Experts & 37 & 39 & 76 \\ \hline
Experts & 31 & 26 & 57 \\ \hline
Total & 68 & 65 & 133 \\ \hline
\end{tabular}
\label{table:study-distribution}
\end{center}
\end{table}
\vspace{-1em}

As observed from Table \ref{table:study-distribution}, the total number of participants were 133, where 57 claimed they are experts and the rest claimed they are not. On the other hand, the number of participants per chart group is almost equal, 68 for conventional and 65 for novel. This result is expected because the chance of being inserted into either charts group is equiprobable (50\% chance), which is hoped to result in a fair comparison between the two chart groups.

By analyzing the incoming data from the 4\textsuperscript{th} qualitative question mentioned earlier, Fig. \ref{fig:data-amount} shows participants' response to the amount of data each chart contains.

\begin{figure}[!ht]
\centering
\includegraphics[trim={0.8in 0.7in 0.8in 0.7in}, width=1\linewidth]{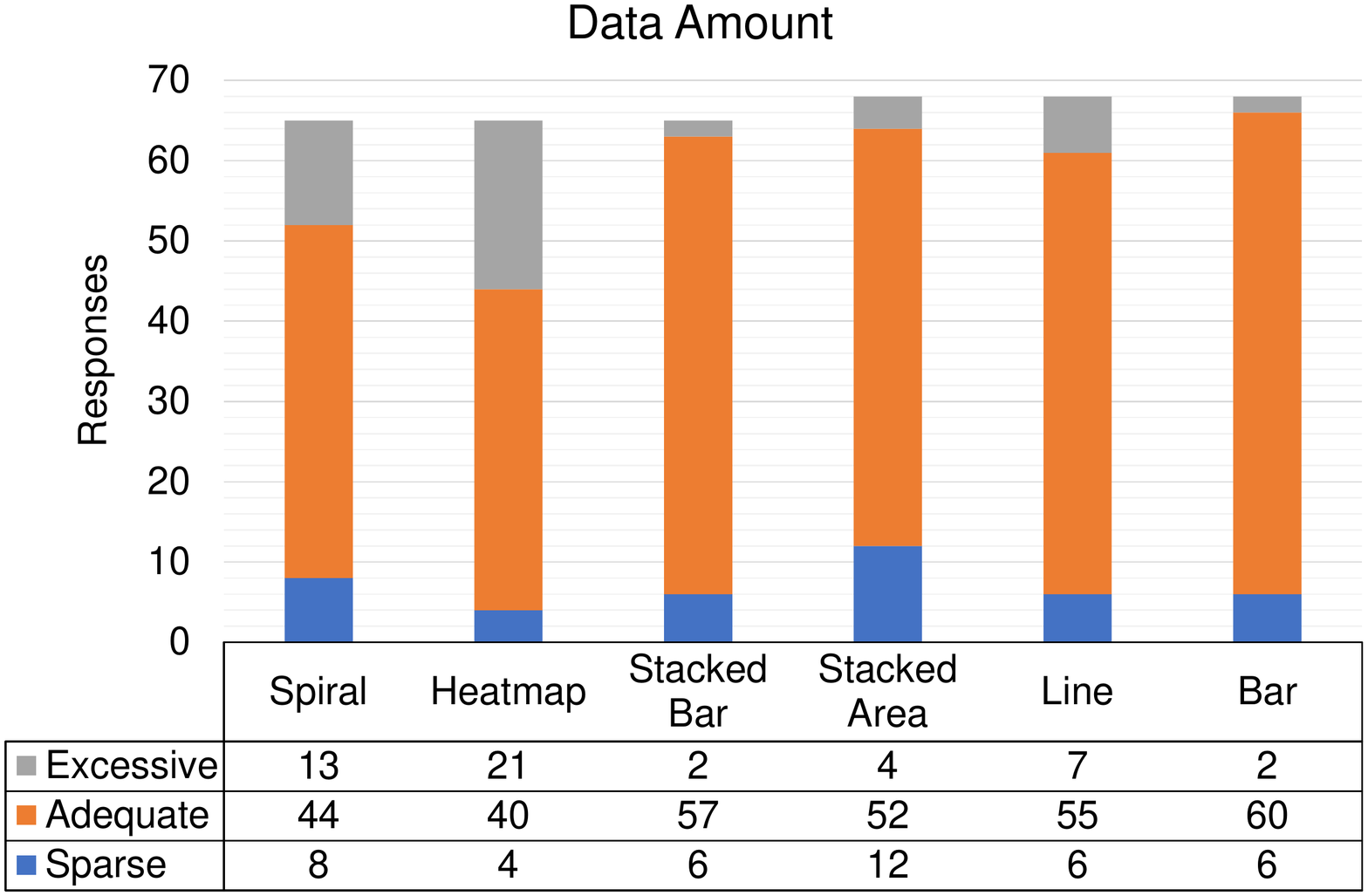}
\caption{Participants' opinion on data amount per chart}
\vspace{-0.75em}
\label{fig:data-amount}
\end{figure}

Each participant had to answer the same qualitative questions about the three charts. In terms of data amount, the heatmap is voted by 21 participants to have an excessive amount of data, while, spiral placed second with 13 votes and stacked bar placed third with 2 votes only. The stacked bar chart is voted, by far, to have an adequate amount of data by most participants among the novel group. On the other hand, the conventional group had quite similar votes to each other, but the bar chart is also favored among that group in terms of data amount. However, stacked area is voted to have the lowest information density among that conventional charts group.

A possible explanation behind conventional charts scoring higher result in the data amount question (voted to have a more adequate amount of data compared to novel charts) is due to participants' familiarity with these charts from different fields. Hence, if participants were more exposed to the novel charts, perception of the amount of data presented in novel group would differ.

The results of the remaining 6 questions (3 qualitative and 3 quantitative) are summarized in Fig. \ref{fig:evaluation-metrics}. The answers to the 3 quantitative questions are evaluated for each chart, normalized and summed under a single metric called analysis questions metric. The maximum percentage each chart can have is 33\%, which is the equivalent of having all participants answering all questions correctly. Moreover, since the qualitative questions were ordinal data ranging between 1 and 5, participants' answers were normalized through dividing by 5 and then also dividing by 3 to normalize the total over the three charts per group. It is worth mentioning that the numbers shown in Fig. \ref{fig:evaluation-metrics} are rounded-off. 

\vspace{-0.5em}
\begin{figure}[!ht]
\centering
\includegraphics[trim={2.5in 0.7in 3in 0.7in}, width=0.8\linewidth]{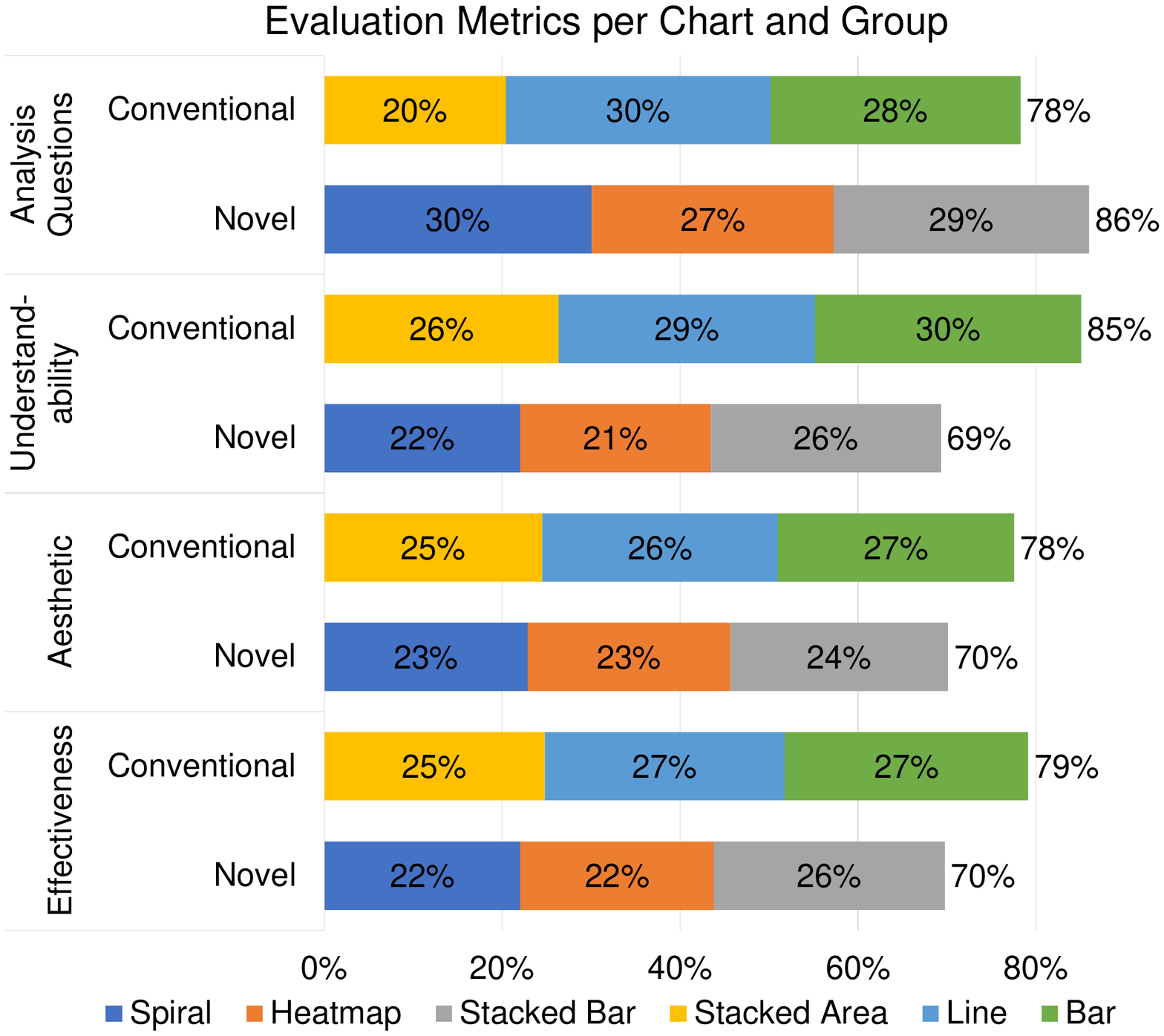}
\caption{Participants' evaluation per chart and group}
\vspace{-1em}
\label{fig:evaluation-metrics}
\end{figure}

Fig. \ref{fig:evaluation-metrics} shows how all the participants rated each chart and how each chart contributes to the total of the group it belongs to. As observed, conventional charts are leading in terms of the qualitative questions (understandability, aesthetics and effectiveness) which are subjective in their nature. In other words, the conventional group is more favored by the participants. However, according to the analysis questions, that are objective in nature, the novel group scored higher by 8\%, which is a large difference in understanding the charts. It can be deduced that even if some charts are simple in the way they look, they can be misleading to the public, which stacked area chart in the conventional group is showing when it dragged down the performance of the whole group. On a side note, scrolling the chart (to the left and the right) might have been missed by some participants, especially in the conventional group, which could be the cause to some of the wrong answers. Consequently, it is an important design aspect that needs to be considered carefully. 

Within the conventional group, the bar chart is superior in all qualitative aspects among its peers. Second comes the line chart which is showing identical energy consumption data, however, it surpasses the bar chart in the analysis questions metric. Third comes the stacked area chart, which is the least favored and the most ambiguous. The order is clear when ranking the conventional charts; however, it is not that clear for the novel group. As observed from Fig. \ref{fig:evaluation-metrics}, the heatmap is the least favored due to the excessive amount of data it presents. Also, it might be the most confusing for the participants as they were not able to interact with it in the right way. Spiral chart, on the other hand, had the highest score in analysis questions. However, it comes in second place after the stacked bar chart in terms of qualitative questions.

Also, noteworthy finding is that, even though participants did not see the other group of charts, both groups favored the bar-like charts, which indicates how familiar and easy-to-understand this type of charts is. Not only that, but they also have an quite similar scores for the data amount metric.

\textbf{So, which group to choose? and shall the less desired charts be discarded or modified?} As presented in Fig. \ref{fig:data-amount} and Fig. \ref{fig:evaluation-metrics}, the decision is not straight-forward, and no group should be discarded completely. However, a decision can be made per chart. The safest action for the stacked area chart is to be discarded or carefully redesigned, where it is the most confusing and the least favored. However, for other charts such as the heatmap, improvements can be applied to compress the amount of data presented without loss of information (e.g. showing energy consumption data in periods of half hours instead of 10 minutes periods). Some charts can be used as is such as the stacked bar chart with appliances from the novel group and the normal bar chart from the conventional group.
\balance
\section{Conclusions}\label{sec5}
To conclude, this paper advocates that energy consumption can be reduced using micro-moments powered by data visualization tools to induce long-term behavioral change. Suitable data visualizations with good recommendations are the perfect tools to raise awareness. A questionnaire is built in a mobile application to evaluate different groups of visualizations. The results show no straight-forward answer as the charts did not convey the same information. Thus, no superiority of one's group over the other is detected, but an action can be taken per chart based on the gathered responses. Some of the limitations were incurred by the fact that React Native is still not well-established in the field of data visualizations. While native mobile development tools have richer libraries to be utilized for that goal, this in hand lengthened the development phase. Another limitation would be that the number of responses received from both genders were not equal (112 males and 21 females). For future work, participants' feedback will be assessed and incorporated into the visualizations. Moreover, implementation of ``what-if'' analysis on these visualizations is also considered. The goal would be to allow consumers to ``see'' for themselves the effects of reducing their energy consumption in a specific aspect and its contribution to the total energy consumption.

\section*{Acknowledgment}
This paper was made possible by National Priorities Research Program (NPRP) grant No. 10-0130-170288 from the
Qatar National Research Fund (a member of Qatar Foundation). The statements made herein are solely the responsibility
of the authors.
\bibliographystyle{IEEEtran}

\bibliography{IEEEabrv.bib,main.bib}{}
\end{document}